\documentstyle[12pt,epsf]{article}
\begin{document}

\newcommand{\nc}{\newcommand}

\def\li{{\rm Li_2}}
\def\Ll{{\rm L}}
\def\eps{\epsilon}
\def\tree{{\rm tree}}
\def\lloop{{\rm loop}}
\def\ib{{\; \bar\imath}}
\def\susy{{\rm SUSY}}
\def\Atree{A^{\rm tree}}
\def\cg{c_\Gamma}
\def\scalar{{\rm scalar}}
\def\tree{{\rm tree}}
\def\subl{{\rm sl}}
\def\si{\sigma}
\def\lr{\leftrightarrow}
\def\la{\langle}
\def\ra{\rangle}
\def\MSbar{$\overline{\rm MS}$}
\def\DRbar{$\overline{\rm DR}$}
\def\spa#1.#2{\left\langle#1 \hskip .15 mm #2\right\rangle}
\def\spb#1.#2{\left[#1 \hskip .15 mm #2\right]}
\def\spaa#1.#2.#3{\langle\mskip-1mu{#1} 
                  | #2 | {#3}\mskip-1mu\rangle}
\def\spbb#1.#2.#3{[\mskip-1mu{#1}
                  | #2 | {#3}\mskip-1mu]}
\def\spab#1.#2.#3{\langle\mskip-1mu{#1} 
                  | #2 | {#3}\mskip-1mu\rangle}
\def\spba#1.#2.#3{\langle\mskip-1mu{#1}^+ 
                  | #2 | {#3}^+\mskip-1mu\rangle}

\def\ellb{{\bar \ell}}
\def\eb{{\bar e}}
\def\pb{{\bar p}}
\def\qb{{\bar q}}
\def\Qb{{\bar Q}}
\def\ub{{\bar u}}
\def\db{{\bar d}}
\def\nub{{\bar \nu}}
\def\Lsnew{\mathop{\rm \widetilde {Ls}}\nolimits}
\def\Ls{\mathop{\rm Ls}\nolimits}
\def\d#1#2{\delta_{#1 #2}}
\def\del{\Delta}
\def\delt{\tilde \Delta}
\def\rtdelta{\sqrt{\Delta_3}}
\def\rtmdelta{\sqrt{-\Delta_3}}
\def\q{{\vphantom{\qb}{q}}}
\def\Q{{\vphantom{\Qb}{Q}}}
\def\msq{m_t^2}
\def\flip#1{\hbox{flip}_{#1}}
\def\exch#1{\hbox{exch}_{#1}}
\def\sstw{\sin^2\theta_W}
\def\Ta{T}
\def\logcoeff{L_{34/12}}

%%%%%%%%%%%%%%%%%%%%%%%%%%%%%%%%%%%%%%%%%%%%%%%%%%%%%%%%%%
%       the stuff below defines \eqalign and \eqalignno in such a
%       way that they will run on Latex
\newskip\humongous \humongous=0pt plus 1000pt minus 1000pt
\def\caja{\mathsurround=0pt}
\def\eqalign#1{\,\vcenter{\openup1\jot \caja
        \ialign{\strut \hfil$\displaystyle{##}$&$
        \displaystyle{{}##}$\hfil\crcr#1\crcr}}\,}
\newif\ifdtup
\def\panorama{\global\dtuptrue \openup1\jot \caja
        \everycr{\noalign{\ifdtup \global\dtupfalse
        \vskip-\lineskiplimit \vskip\normallineskiplimit
        \else \penalty\interdisplaylinepenalty \fi}}}
\def\eqalignno#1{\panorama \tabskip=\humongous
        \halign to\displaywidth{\hfil$\displaystyle{##}$
        \tabskip=0pt&$\displaystyle{{}##}$\hfil
        \tabskip=\humongous&\llap{$##$}\tabskip=0pt
        \crcr#1\crcr}}

%%%%%%%%%%%%%%%%%%%%%%%%%
% Equation labeling defined to make numbering by \equn
% and labeling works with \label{TheLabel} as per normal LaTeX
%
% have section numbered equations
\def\theequation{\thesection.\arabic{equation}}
%a.s. \def\theequation{\thesection.\arabic{eqnumber}}

\newcounter{eqnumber}
\renewcommand{\theeqnumber}{\arabic{eqnumber}}
\def\equn{
\refstepcounter{eqnumber}
%\eqno({\rm \theeqnumber})
\eqno({\rm \theequation})
 }

%%%%%%%%%%%%%%%%%%%%%%%%%%%%%%%%%%%%%%%%%%%%%%%%%%%%%%%%%%%%%%%%%%%%%%%%%%
% Some macros for mathematics

\newcommand{\gtrsim}{\raisebox{.2em}{$\rlap{\raisebox{-.5em}{$\;\sim$}}>\,$}}
% some definitions
\def\eqn#1{eq.~(\ref{#1})}
\def\Eqn#1{Eq.~(\ref{#1})}
\def\eqns#1{eqs.~#1}
\def\Eqns#1{Eqs.~#1}
\def\sec#1{section~{\ref{#1}}}
\def\Sec#1{Section~{\ref{#1}}}
\def\app#1{appendix~\ref{#1}}
\def\App#1{Appendix~\ref{#1}}
\nc{\fig}[1]{Fig.~\ref{fig:#1}}
\nc{\figs}[2]{Figs.~\ref{fig:#1} and \ref{fig:#2}}

\renewcommand{\slash}[1]{/\kern-7pt #1}
\newcommand{\beq}{\begin{equation}}
\newcommand{\eeq}{\end{equation}}
\newcommand{\beqn}{\begin{eqnarray}}
\newcommand{\eeqn}{\end{eqnarray}}
\newcommand{\beqns}{\begin{eqnarray*}}
\newcommand{\eeqns}{\end{eqnarray*}}
\newcommand{\nn}{\nonumber}
 
\def\cA{{\cal A}}
\def\cB{{\cal B}}
\def\cC{{\cal C}}
\def\cD{{\cal D}}
\def\cE{{\cal E}}
\def\cF{{\cal F}}
\def\cG{{\cal G}}
\def\cH{{\cal H}}
\def\cI{{\cal I}}
\def\cK{{\cal K}}
\def\cM{{\cal M}}
\def\cN{{\cal N}}
\def\cO{{\cal O}}
\def\cP{{\cal P}}
\def\cR{{\cal R}}
\def\cS{{\cal S}}
\def\cX{{\cal X}}
\def\cZ{{\cal Z}}
\def\to{\rightarrow}
\newcommand{\lra}{\leftrightarrow}
\newcommand{\as}{\alpha_s}
 
\newcommand{\Nf}{N_f}
\newcommand{\Nc}{N_c}
 
%%%%%%%%%%%%%%%%%%%%%%%%%%%%%%%%%%%%%%%%%%%%
 
\begin{titlepage}
\begin{flushright}
SLAC--PUB--7750\\
ETH-TH/98-07\\
CERN-TH/98-66 \\
hep-ph/9803250\\
March 1998
\end{flushright}
\begin{center}
\vspace*{2.5cm}
{\Large\bf Helicity Amplitudes for ${\cal O}(\alpha_s)$ Production of 
 $W^+ W^-,  W^\pm Z, Z Z, W^\pm \gamma,$ or $Z \gamma$ Pairs 
 at Hadron Colliders} 
\vskip 0.5cm
{\large L. Dixon$^{1,a}$, Z. Kunszt$^2$ and A. Signer$^3$} \\
\vskip 0.4cm
{$^1$\it Stanford Linear Accelerator Center, Stanford University, 
Stanford, CA 94309, USA} \\
\vskip 0.1cm
{$^2$\it Theoretical Physics, ETH Z\"urich, Switzerland} \\
\vskip 0.1cm
{$^3$\it Theory Division, CERN, CH-1211 Geneva 23, Switzerland} \\
 
\vskip 4cm
\end{center}
 
\begin{abstract}
\noindent
We present the one-loop QCD corrections to the helicity
amplitudes for the processes 
$q\qb \to W^+ W^-,\ Z \, Z,\ W^\pm Z, \ W^\pm \gamma$, or $Z \gamma$,
including the
subsequent decay of each massive vector boson into a pair of leptons.
We also give the corresponding tree-level amplitudes with an additional 
gluon radiated off the quark line.  Together, these amplitudes provide all
the necessary input for the calculation of the next-to-leading order QCD 
corrections to the production of any electroweak vector boson pair at hadron 
colliders, including the full spin and decay angle correlations. 
\end{abstract}

\vskip 1cm
\begin{center}
{\sl Submitted to Nuclear Physics B}
\end{center}
 
\vfill
\noindent\hrule width 3.6in\hfil\break
${}^{a}${\tenrm Research supported by the US Department of
Energy under grant DE-AC03-76SF00515.}\hfil\break

\end{titlepage}
%%%%%%%%%%
\setcounter{footnote}{0}
 
\newpage
 
%%%%%%%%%%%%%%%%%%%%%%%%%%%%%%%%%%%%%%%%%%%%%%%%%%%%%%%%

\section{Introduction}

With the increasing energy available at hadron colliders, electroweak
gauge boson pair production becomes more and more important.  Both
experimental collaborations at the Tevatron, CDF \cite{CDFref} and D0
\cite{D0ref}, have performed studies of pair production processes such
as $p \pb \to W^+ W^-, Z Z, W^\pm Z,W^\pm \gamma$, or $Z \gamma$.
These studies considered purely leptonic decays of the massive vector
bosons in the pair, as well as decays into two jets plus leptons.
Already, some events have been found above the background, in
accordance with Standard Model predictions. The amount of available
data will increase by roughly a factor of 20 at the upgraded Tevatron,
and by a factor of 1000 once the Large Hadron Collider at CERN (LHC)
starts operating.  A summary of the present experimental situation and
a more complete list of earlier studies can be found in
\cite{CDFD0talks}. 

These processes are quite interesting in several respects.  Most of
all, they can be used to measure the vector boson trilinear couplings
predicted by the Standard Model \cite{HWZTriple}.  Any kind of
anomalous couplings, or decays of new particles into vector boson
pairs, would result in deviations from these predictions. In
particular, if the Higgs boson is heavy enough it will decay mainly
into $W^+ W^-$ and $Z Z $ pairs \cite{EHLQ}.  Thus, the search
for Higgs bosons in the few hundred GeV mass range is intimately
connected to pair production of vector bosons.  A detailed
understanding of these Standard Model processes is therefore
mandatory.

Due to its importance, hadronic pair production of electroweak vector
bosons has received a lot of attention in the literature.  The tree-level
cross-sections for the hadronic production of $W^+ W^-, \ Z Z, \ W^\pm Z$,
as well as $W^\pm \gamma$ and  $Z \gamma$ pairs were computed long ago
\cite{TreePair}.  The one-loop (${\cal O}(\alpha_s)$) QCD corrections to 
these cross-sections have been computed in \cite{ZZOhn,ZZit} for $Z
Z$, in \cite{WZOhn,WZit} for $W^\pm Z$, in \cite{WWOhn,WWit} for $W^+
W^-$, and in \cite{STN,yOhn} for $W^\pm \gamma$ and $Z \gamma$.  These
computations were all done with the traditional method of evaluating
directly the squared amplitude through interference (cut) diagrams and
evaluating the traces in $D = 4-2\eps$ dimensions.  As a consequence,
the computed cross-sections were summed over all $W$ and $Z$
polarization states.

A more realistic treatment of these processes can be obtained by
properly including the decay of the vector bosons into massless
fermions.  In fact, vector bosons are identified through these decay
products.  Thus, the comparison of theory and experiment is much
easier, since cuts on the kinematics of the decay products can easily
be added to the computation.  As one example of the importance of
decay-angle correlations, it has been proposed \cite{GOWBBHK,DD} to
search for the Higgs boson at the LHC in the intermediate mass range
$m_H = 155-180$~GeV in the channel $H \to W^+W^- \to \ell^+ \nu \ell^-
\nub$.  To reduce the continuum $W^+W^-$ background, one can exploit
\cite{DD} the anti-correlation between $W$ helicities for the signal
process, as well as the strong correlation between the $W$ helicity
and the decay lepton direction; the signal peaks when $\ell^+$ and
$\ell^-$ are nearly collinear, $\cos\theta_{\ell^+\ell^-} \approx 1$,
while the background is relatively flat in this variable.  For such a
search it is clearly important to understand as well as possible the
background distributions for $\cos\theta_{\ell^+\ell^-}$ and other
kinematic variables.

In the narrow-width approximation, vector boson decay is simple to
implement at the amplitude level.  Because the couplings of vector bosons
to fermions are spin-dependent (especially the purely left-handed $W$
couplings), it is natural to employ the helicity method and compute
amplitudes for massless external states of definite helicity.  The
tree-level helicity amplitudes for massive vector-boson pair production with
subsequent decay into leptons were first computed in \cite{GK}.  The
authors of \cite{GK} also showed that the effects of decay-angle
correlations are significant.

In \cite{OhnSpin}, the above two approaches were merged to get a 
more complete next-to-leading order treatment of vector boson pair
production.  In this work spin correlations were included everywhere except
for the virtual contribution.  Furthermore, the calculations were extended
to include also non-standard triple-vector-boson couplings.  However, only
numerical results were presented, making it difficult to use these
results in future computations. 

The present paper closes this gap by presenting all helicity amplitudes 
required for next-to-leading order (in the strong coupling $\alpha_s$) 
hadronic production of a vector boson pair.  In particular, we 
give the one-loop amplitudes with a virtual gluon for the processes
\beqn
\qb q \to & \!\! W^- W^+ & \!\! \to   
      (\ell\, \nub)\ +\ (\ellb^\prime \, \nu^\prime) \,, 
  \label{procWW} \\
\qb q \to & \!\! Z \ Z & \!\! \to   
  (\ell \, \ellb)\ +\ (\ellb^\prime \, \ell^\prime) \,,
  \label{procZZ}  \\
\ub d \to & \!\!  W^- Z & \!\! \to 
      (\ell \, \nub)\ +\ (\ellb^\prime \, \ell^\prime) \,,
  \label{procWZ} \\
\ub d \to & \!\!  W^- \gamma & \!\! \to 
      (\ell \, \nub)\ +\ \gamma \,,
  \label{procWgamma} \\
\qb q \to & \!\! Z \ \gamma & \!\! \to   
  (\ell \, \ellb)\ +\ \gamma \,.
  \label{procZgamma}  
\eeqn
The processes with a $W^+Z$ or a $W^+\gamma$ pair as intermediate state
can be obtained from \eqn{procWZ} and \eqn{procWgamma} by a CP
transformation.  The decay of the vector bosons 
into leptons is included in the narrow-width approximation.
We also present the tree-level amplitudes for the same processes with an
additional gluon radiated off the quark line.  (The gluon may also appear
in the initial state. The corresponding amplitudes can be obtained by
crossing symmetry.)  Both sets of amplitudes 
are needed for a complete next-to-leading order computation of the 
cross-sections for vector-boson pair production.  

In the spinor helicity formalism \cite{SpinorHelicity},
the tree-level amplitudes are trivial to obtain and the results are 
very compact.   For the one-loop amplitudes, it was not
necessary to do a full computation.  Almost all of the terms could be 
extracted from one of the helicity amplitudes for the process 
$e^+ e^- \to q\qb Q\Qb$ as presented in \cite{BDKW4q}, where $q$ and $Q$
are massless quarks of different flavor.      
In fact, knowledge of the `primitive amplitude' (see section \ref{PriAmp})
for the subleading-in-color piece of $e^+ e^- \to q\qb Q\Qb$
is sufficient to obtain all one-loop amplitudes for the processes listed in 
eqs.~(\ref{procWW})--(\ref{procWZ}). The amplitude for the process
(\ref{procWgamma}) can also be obtained without doing a full computation. It
is sufficient to replace the $Z$ in the process (\ref{procWZ}) by a virtual
photon $\gamma^*$. Then the desired amplitude can be extracted from the
collinear limit of the decay products of the virtual photon. Finally,
these results allow for the construction of the amplitude for the last
missing process, \eqn{procZgamma}. This last result could also be obtained
from the known subleading-in-color primitive amplitude for 
$e^+ e^- \to q \qb g$ \cite{GieleGlover,BDK2q2g}.

Besides their contribution to next-to-leading order pair production rates,
the one-loop amplitudes also contribute, via their absorptive parts, to
kinematic structures which are odd under `naive' time reversal 
(the reversal of all momentum and spin vectors in a process).
Such terms are not present at tree level; also, they are washed out if
one integrates over all the leptonic decay angles.  Analogous effects
were considered quite some time ago, in the production of $W\,+\,$1 jet 
at hadron colliders \cite{HHKTodd}.

In section \ref{PriAmp} we exploit the results of \cite{BDKW4q} 
in order to obtain certain primitive amplitudes, called $A^a$ and $A^b$, 
which serve as building blocks for the construction of all the 
one-loop amplitudes for 
$q\qb \to W^+ W^-, \ Z Z, \ W^\pm Z \to $ 4 leptons.
This construction will be performed in section \ref{dressing}. Section
\ref{photons} is devoted to the processes with a real photon in the final
state. Finally, in section \ref{conclusion} we compare our results to the 
literature \cite{ZZit,WZit,WWit,STN} and present our conclusions. 

%%%%%%%%%%%%%%%%%%%%%

\section{Primitive amplitudes for $W^+ W^-,\ Z \, Z,\ W^\pm Z $
         \label{PriAmp} }
\setcounter{equation}{0}

\subsection{Preliminaries \label{preliminaries}}

It is by now standard to use the helicity method and color ordering of the
amplitudes to simplify one-loop calculations in QCD (for a review see
e.g. \cite{OneLoopHel}). The results for the helicity amplitudes will be
expressed in terms of spinor inner products, 
\beq
\spa i.j \equiv \la k_i^- | k_j^+ \ra, \quad 
\spb i.j  \equiv \la k_i^+ | k_j^- \ra ,
\eeq
where $|k_i^\pm \ra$ is the Weyl spinor for a massless particle with
momentum $k_i$.  The spinor inner products are antisymmetric and satisfy 
$\spa i.j \spb j.i = 2 k_i \cdot k_j \equiv s_{ij}$. For later use we also
define
\beqn
\label{defsAB}
\la i|l|j \ra &\equiv& \la k_i^- |\slash{k_l}| k_j^- \ra ,\nn \\
\la i|lm|j \ra &\equiv& \la k_i^- |\slash{k_l}\,\slash{k_m}| k_j^+ \ra ,\nn \\
% \left[ i|l|j \right] &\equiv& \la k_i^+ |\slash{k_l}| k_j^- \ra ,\nn \\
\la i|(l+m)|j \ra &\equiv& 
    \la k_i^- |(\slash{k_l}+\slash{k_m})| k_j^- \ra ,  \\
% \left[ i|(l+m)|j \right] &\equiv& 
%   \la k_i^+ |(\slash{k_l}+\slash{k_m})| k_j^+ \ra ,  \\
\la i|(l+m)(n+r)|j \ra &\equiv& 
\la k_i^- |(\slash{k_l}+\slash{k_m})(\slash{k_n}+\slash{k_r})| k_j^+ \ra ,
 \nn \\
% \left[ i|(l+m)(n+r)|j \right]  &\equiv&
% \la k_i^+ |(\slash{k_l}+\slash{k_m})(\slash{k_n}+\slash{k_r})| k_j^- \ra
% , \nn
\left[ i| \ldots |j\right]  &\equiv& \la i|\ldots |j\ra 
           \bigg|_{k_{\{i,j\}}^\pm \to k_{\{i,j\}}^\mp} \nn
\eeqn
and
\beqn
\label{defsS}
s_{ij} &\equiv& (k_i + k_j)^2 , \qquad t_{ijm} \equiv (k_i + k_j + k_m)^2, 
  \nn \\
\delta_{12} &\equiv& s_{12} - s_{34} - s_{56} , \qquad
\delta_{34} \equiv s_{34} - s_{12} - s_{56} , \qquad
\delta_{56} \equiv s_{56} - s_{12} - s_{34} , 
  \\
\Delta_3 &\equiv& s_{12}^2 + s_{34}^2 + s_{56}^2 - 2 s_{12} s_{34} - 
  2 s_{12} s_{56} - 2 s_{34} s_{56}. \nn
\eeqn

For all the processes $q\qb \to V_1 V_2 \to $ 4 leptons, 
where $V_{1,2} \equiv W^\pm$ or $Z$, the color ordering of the amplitudes 
is trivial:  All diagrams have the same color factor.  
Thus the full amplitude can be obtained by taking the one
and only subamplitude (partial amplitude) and multiplying it by the
overall color factor. 

%%%%%%%%%%%%%%%%%%%%%%%%%%%%%%%%%%%%%%%%%%%%%%%%%%%%%%%%%%%%%%%%%%%%%%%%
\begin{figure}[t]
   \vspace{-3.5cm}
   \epsfysize=24cm
   \epsfxsize=16cm
   \centerline{\epsffile{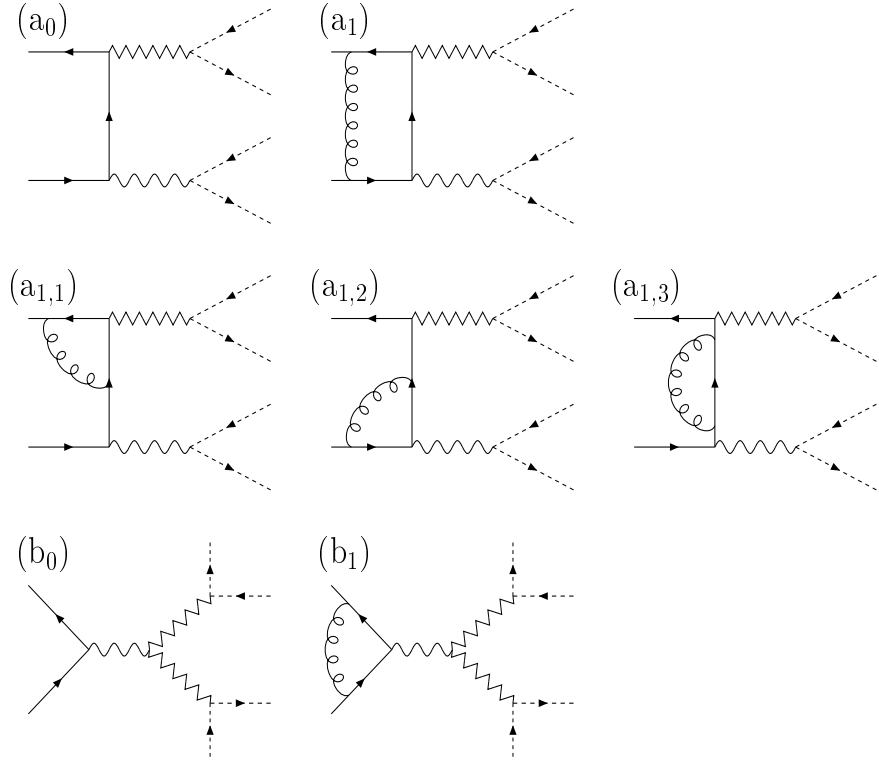}}
   \vspace*{-10.0cm}
\caption[dummy] {\small (a$_0$) box-parent tree graph; (a$_1$) box-parent
   one-loop graph;  (a$_{1,1}$) -- (a$_{1,3}$) additional one-loop graphs
   obtained from the box-parent graph; (b$_0$) triangle-parent tree graph;
   (b$_1$) triangle-parent one-loop graph. Solid (dashed) lines represent
   quarks (leptons). In order to distinguish the possibly different
   vector bosons we used zigzag and wavy lines. For the graphs b$_{0}$ and
   b$_{1}$ we show the diagrams for the $W^+ W^-$ intermediate state. 
   \label{fig:diag1} }
\end{figure}
%%%%%%%%%%%%%%%%%%%%%%%%%%%%%%%%%%%%%%%%%%%%%%%%%%%%%%%%%%%%%%%%%%%%%%%

Still, the diagrams which contribute to
$q\qb \to V_1 V_2 \to $ 4 leptons
naturally fall into two classes, which can be distinguished by their
different dependence on the electroweak coupling constants, and which
therefore are separately gauge invariant.  (See \fig{diag1}.) 
This decomposition is analogous to the decomposition of QCD amplitudes into
primitive amplitudes \cite{BDK2q3g}.  The first class (diagrams (a$_i$) of
\fig{diag1}) has no triple-electroweak-vector-boson vertex. 
There is only one such tree graph, (a$_0$).
We call the one-loop box graph (a$_1$) the `box-parent' because the other
one-loop graphs in this class, (a$_{1,1}$) -- (a$_{1,3}$), can be obtained
from (a$_1$) by sliding the virtual gluon line
around, while leaving the electroweak vector bosons fixed.
(Bubble graphs where a gluon dresses a massless external quark line are 
zero in dimensional regularization.)
The tree-level primitive amplitude $A^{\tree,a}$ is defined to be
just the contribution of the graph (a$_0$), omitting all coupling constant
prefactors.  The one-loop primitive amplitude $A^{a}$ is similarly defined
by the sum of graphs (a$_1$) and (a$_{1,1}$) -- (a$_{1,3}$).
The second class (diagrams (b$_i$) of \fig{diag1}) contains a 
three-boson vertex.  In this class, there are no further graphs
besides the parent graphs.  These primitive amplitudes will be 
denoted by $A^{\tree,b}$ and $A^b$.  Because $A^b$ consists of a single
triangle diagram, it is much simpler than $A^a$. 

We present amplitudes in the dimensional reduction \cite{Siegel} 
or four-dimensional helicity \cite{StringBased} variants of dimensional
regularization, which are equivalent at one loop.  The conversion to
other variants is straightforward \cite{KunsztFourPoint}.

%%%%%%%%%%%%%%%%%%

\subsection{Relation to subleading-color 
$e^+e^- \to q\qb Q\Qb$ amplitude \label{relation}}

\begin{figure}[t]
   \vspace{-3.5cm}
   \epsfysize=27cm
   \epsfxsize=18cm
   \centerline{\epsffile{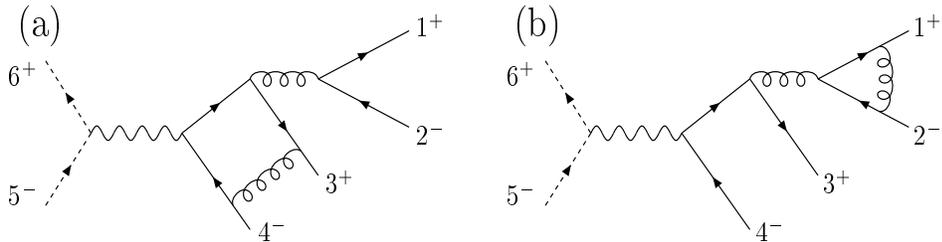}}
   \vspace*{-18.8cm}
\caption[dummy] {\small Parent diagrams for $A_6^\subl(1,2,3,4)$, i.e. the
   subleading-in-color part of $e^+ e^- \to q \qb Q \Qb$. Solid (dotted)
   lines represent quarks (electrons). Note that the labeling used in
   this figure and in section \ref{relation} corresponds to the one
   used in ref. \cite{BDKW4q}.  
   \label{fig:diag2} }
\end{figure}

The one-loop amplitudes for 
$q\qb \to V_1 V_2 \to $ 4 leptons 
can be obtained {\it almost} completely from the subleading-color amplitude
$A_6^{\subl}(1,2,3,4)$ for the process $e^+e^- \to q\qb Q\Qb$,
as given in eqs.~(3.16)--(3.21) of
\cite{BDKW4q}\footnote{In order to simplify the discussion at this point,
  we use the labeling of \cite{BDKW4q} in section
  \ref{relation}. The labeling used in the rest of this
  paper can be obtained from it by the replacement 
  $\{1,2,3,4,5,6\} \to \{5,6,2,1,3,4\}$.}.
The `parent' graphs for that amplitude are shown in \fig{diag2}.
Graph (a) is obviously very similar to corresponding box-parent graph (a$_1$) 
in \fig{diag1}.   The quark pair $\{1,2\}$ can be replaced by a lepton pair,
and the virtual gluon that connects the two different quark lines can be 
replaced by an electroweak vector boson.  The fact that the vector boson 
can have axial-vector as well as vector couplings, while the gluon has 
only vector couplings, can be accounted for in the helicity formalism 
simply by dressing the graphs with the appropriate left- and right-handed 
couplings of fermions to electroweak bosons.

If $A_6^{\subl}(1,2,3,4)$ were truly a primitive amplitude, then
one would be able to get $q\qb \to V_1 V_2 \to$ 4 leptons from it with no 
further information.  However, the $A_6^{\subl}(1,2,3,4)$ as given in
\cite{BDKW4q} is not truly primitive, for two reasons.

Firstly, the diagrams (a) and (b) in \fig{diag2} are combined together
in $A_6^{\subl}(1,2,3,4)$.  However, it is no problem to remove diagram (b),
because its contribution has been identified explicitly in \cite{BDKW4q} 
(as the terms in the second bracket in eq.~(3.19) for $V^\subl$).  
Both diagrams (b$_1$) of \fig{diag1} and (b) of \fig{diag2} are QCD 
vertex corrections, which evaluate to a universal factor times
the corresponding tree graph.
Thus the terms from diagram (b) in \fig{diag2}, which appeared multiplied 
by $A_6^{\tree,\subl}$ in \cite{BDKW4q}, will reappear here in $A^b$, 
multiplied instead by $A^{\tree,b}$.

%%%%%%%%%%%%%%%%%%%%%%%%%%%%%%%%%%%%%%%%%%%%%%%%%%%%%%%%%%%%%%%%%%%%%%%%
%%%%%%%%%%%%%%%%%%%%%%%%%%%%%%%%%%%%%%%%%%%%%%%%%%%%%%%%%%%%%%%%%%%%%%%

Secondly, and more non-trivially, another class of graphs has also been
included in $A_6^\subl(1,2,3,4)$ in addition to those shown in \fig{diag2}
--- those where the quark pair $\{5,6\}$ and the lepton pair $\{1,2\}$ are 
exchanged.  In the process $e^+e^- \to q\qb Q\Qb$, both classes of graphs 
have equal weight, but in $q\qb \to V_1 V_2 \to$ 4 leptons they do not 
(except for the $ZZ$ case).  The permutation that generates the 
`exchange' terms is
\beq
\hbox{exchange:} \hskip 1 cm 
1 \leftrightarrow 6 \,, \hskip 1 cm  
2 \leftrightarrow 5 \,.
\label{ExchangeSymmetry}
\eeq
We need to delete the `exchange' terms from the (a) terms in 
$A_6^\subl(1,2,3,4)$.  Since `exchange' takes $t_{123}$ to $t_{124}$,
part of this deletion is easy to implement --- we want to omit 
all terms with a cut in the $t_{124}$ channel, for example.
Also, the second term in the tree amplitude $A_6^{\tree,\subl}$
in eq.~(3.16) of \cite{BDKW4q} should clearly be omitted.
But this is not enough.  There is still a potential ambiguity about
which terms come from the desired graphs, and which terms from 
the exchange graphs, in the coefficients of 
$\ln(s_{12})$, $\ln(s_{56})$, $\ln(s_{34})$ and the three-mass triangle 
integral $I_3^{3{\rm m}}(s_{12},s_{34},s_{56})$, 
as well as the rational-function terms.

Fortunately, it turns out that all of the ambiguous terms, except for
the rational-function terms, {\it were} given in eqs.~(3.20) and (3.21) of
\cite{BDKW4q} as the contributions of the desired graphs,
not including the exchange.  The exchange that was desired for 
$e^+e^- \to q\qb Q\Qb$ was then added explicitly, at the end of
eq.~(3.20). This same statement is {\it not} true for a second (simpler)
form of $A_6^{\subl}(1,2,3,4)$, given in eqs.~(12.9)--(12.12) of
\cite{BDK2q2g}. In that form, the exchange symmetry was used extensively
to simplify the coefficients of $\ln(s_{12})$ and $I_3^{3{\rm m}}$, at the 
cost, however, of hopelessly entangling the desired graphs with their
exchange.  (However, we have managed to significantly simplify even the
non-exchange form of these coefficients, as one may appreciate by
comparing \eqn{TaWW} below to eq.~(3.21) of \cite{BDKW4q}.)

We have calculated the desired rational-function terms.
We find that $F^\subl(1,2,3,4)$ as given in eq.~(3.20) of \cite{BDKW4q}
is also valid for the non-exchange case, provided only that the 
final `$+$exchange' instruction is dropped, and that the following
rational-function terms are added to the existing terms:
\beq
\delta p \equiv 
-{1\over2}{1\over\spab3.{(1+2)}.4} \left(
  {\spb1.6^2\over \spb1.2\spb5.6} + {\spa2.5^2\over\spa1.2\spa5.6} \right)
- {1\over2} {\spab2.{(1+3)}.6\spab5.{(2+4)}.1
     \over s_{12} \, s_{56} \, \spab3.{(1+2)}.4} \,. 
\label{deltap}
\eeq

%%%%%%%%%%%%%%%%%%%

\subsection{Virtual primitive amplitudes}

After simplifying the coefficients of $\ln(s_{12})$ and $I_3^{3{\rm m}}$,
we can write a fairly compact form for the required one-loop primitive
amplitudes, 
\beq
A^\alpha(1,2,3,4,5,6), \qquad \alpha=a,b.
\label{PrimAmpDef1}
\eeq
In the application to the processes $q\qb \to V_1 V_2 \to$ 4 leptons,
the quark and anti-quark will always be drawn from the set $\{1,2\}$,
while the leptonic decay products of a vector boson will be $\{3,4\}$
or $\{5,6\}$.  However, the exact correspondence depends on the fermion 
helicity; i.e., the same function will appear with various permutations 
of its arguments $(1,2,3,4,5,6)$ in section~\ref{dressing}.
The helicity assignments in \eqn{PrimAmpDef1} are 
$(1^-,2^+,3^-,4^+,5^+,6^-)$ (when all particles are considered outgoing).
For the rest of this section the arguments $(1,2,3,4,5,6)$ will be implicit.

First define the flip symmetry,
\beq
\flip{1}: \hskip 8 mm
1\leftrightarrow 2\,, \hskip 0.4 cm  
3\leftrightarrow 5\,, \hskip 0.4 cm  
4\leftrightarrow 6\,, \hskip 0.4 cm 
\spa{a}.{b} \leftrightarrow \spb{a}.{b} \,. 
\label{FlipSymmetry}
\eeq
The box-parent tree graph is given by 
\beq
A^{\tree,a} = 
%%%%% begin WW:  Atreea
i \, { \spa1.3\spb2.5 \spab6.{(2+5)}.4 \over s_{34}\,s_{56}\,t_{134} } 
%%%%% end WW:  Atreea
\,.
\label{treea}
\eeq
The triangle-parent tree graph is given by
\beq
A^{\tree,b} =  
%%%%% begin WW:  Atreeb
{ i \over s_{12}\,s_{34}\,s_{56} }
\Bigl[ \spa1.3\spb2.5 \spab6.{(2+5)}.4
     + \spb2.4\spa1.6 \spab3.{(1+6)}.5 \Bigr]
%%%%% end WW:  Atreeb
\,,
\label{treeb}
\eeq
in a form that agrees with \cite{GK}. 

At one loop, we perform a decomposition into a universal divergent piece $V$,
and finite pieces $F^\alpha$,
\beq
A^{\alpha}\ =\ \cg \Bigl[ 
A^{{\rm tree},\, \alpha } V + i \, F^\alpha \Bigr]\, ,
\label{VFdecomp}
\eeq
where $\alpha = a,b$, the prefactor is
\beq\eqalign{
  \cg\ &=\ {1\over (4\pi)^{2-\eps}}
 {\Gamma(1+\eps)\Gamma^2(1-\eps)\over\Gamma(1-2\eps)}\,, \cr
}\label{cgammadef}
\eeq
and we work in dimensional regularization with $D= 4 - 2 \eps$. 
The amplitudes presented here are ultraviolet-finite; 
all poles in $\eps$ arise from virtual infrared (soft or collinear) 
singularities.

The divergent pieces are given by
\beq
V = 
%%%%% begin WW:  V
  - {1\over \eps^2} \left( {\mu^2 \over -s_{12}}\right) ^\eps 
  - {3\over 2 \eps} \left( {\mu^2 \over -s_{12}}\right) ^\eps 
  - {7\over 2} 
%%%%% end WW:  V
\,.
\label{Vab}
\eeq
The finite piece of the triangle-parent contribution vanishes,
\beq
F^b = 
%%%%% begin WW:  Fb
0
%%%%% end WW:  Fb
\,,
\label{Fb}
\eeq
and the corresponding box-parent contribution is given by
\beqn
F^a & = &
%%%%% begin WW: Fa
\Biggl[ { \spa1.3^2\spb2.5^2 
    \over \spa3.4\spb5.6 \,t_{134}\,\spab1.{(5+6)}.2 } 
 - { \spab2.{(5+6)}.4^2 \spab6.{(2+5)}.1^2
    \over \spb3.4\spa5.6 \,t_{134}\,\spab2.{(5+6)}.1^3 } \Biggr]
     \Lsnew^{2 {\rm m} h}_{-1}(s_{12},t_{134};s_{34},s_{56}) \nn \\
%
% single logs
% vphantom is here in order to insert a dummy variable which preserves the
% parentheses for mathematica, so that `\flip{1}' is handled correctly...
&+& \vphantom{dummy 1}  \Biggl[
  {1\over 2} {\spab6.1.4^2 \,  t_{134}
    \over \spb3.4 \spa5.6 \spab2.{(5+6)}.1}
            {\Ll_1\Bigl( {-s_{34} \over -t_{134}}\Bigr) \over  t_{134}^2 }
+ 2 \, {\spab6.1.4 \spab6.{(2+5)}.4
   \over \spb3.4 \spa5.6 \spab2.{(5+6)}.1 }
             {\Ll_0\Bigl( {-t_{134} \over -s_{34}} \Bigr) \over s_{34} }
  \nn \\
&& \quad
- \ {\spa1.6 \spa2.6 \spb1.4^2 \, t_{134}
   \over \spb3.4 \spa5.6 \spab2.{(5+6)}.1^2}
             {\Ll_0\Bigl({- t_{134} \over -s_{34}} \Bigr) \over  s_{34} }
- {1\over 2} { \spa2.6 \spb1.4 \spab6.{(2+5)}.4
    \over \spb3.4 \spa5.6 \spab2.{(5+6)}.1^2 }
             \ln\biggl({ (-t_{134})(-s_{12}) \over (-s_{34})^2 }\biggr)  
  \nn \\
&& \quad
- \ {3\over 4} {\spab6.{(2+5)}.4^2  
    \over \spb3.4 \spa5.6 \, t_{134} \, \spab2.{(5+6)}.1 }
             \ln\biggl({ (-t_{134})(-s_{12}) \over (-s_{34})^2 }\biggr)
% the [extra] ln(s_34/s_12) terms for one graph
\, + \, \logcoeff \, \ln\Bigl({-s_{34} \over -s_{12} }\Bigl)  
     \, - \, \flip1 \Biggr]  \nn \\
%
% 3 mass triangle and poly terms:
%
&+& \Ta \, I_3^{3 \rm m}(s_{12}, s_{34}, s_{56})
+ {1\over 2} {  ( t_{234} \d12 + 2 s_{34} s_{56})
  \over \spab2.{(5+6)}.1 \Delta_3}
    \biggl( {\spb4.5^2 \over \spb3.4 \spb5.6} 
         + {\spa3.6^2 \over \spa3.4 \spa5.6} \biggr) \nn \\
&+& {\spa3.6 \spb4.5 (t_{134} - t_{234}) \over \spab2.{(5+6)}.1 \Delta_3} 
- {1\over2} {\spab6.{(2+5)}.4^2
     \over  \spb3.4 \spa5.6 t_{134} \spab2.{(5+6)}.1} 
%%%%% end WW: Fa
\,, \label{Fa}
\eeqn
where `$\flip{1}$' is to be applied only to the terms 
inside the brackets ($[\ ]$) in which it appears. 
The coefficient of the logarithm reads
\beq
\eqalign{
\logcoeff & = 
%%%%% begin WW: logcoeff
  {3\over2} { \d56 \, (t_{134}-t_{234}) \, \spab3.{(1+2)}.4 \spab6.{(1+2)}.5 
              \over \spab2.{(5+6)}.1 \Delta_3^2 }
+ {3\over2} {\spa3.6\spbb4.{(1+2)(3+4)}.5 
              \over \spab2.{(5+6)}.1 \, \Delta_3 } \cr
& \hskip 0.3 cm 
+ {1\over2} { \spab3.4.5 \spbb4.{(5+6)(1+2)}.5 
    \over \spb5.6\spab2.{(5+6)}.1 \Delta_3 } 
+ { \spb1.4\spa2.6 \, t_{134} \, 
   \bigl( \spa3.6 \, \d12 - 2 \, \spaa3.{45}.6\bigr)
  \over \spa5.6\spab2.{(5+6)}.1^2 \Delta_3 } \cr
& \hskip 0.3 cm 
+ {1\over2} { t_{134} \over \spab2.{(5+6)}.1 \Delta_3 } 
  \biggl( { \spa3.4\spb4.5^2\over\spb5.6 } 
       + { \spb3.4\spa3.6^2\over\spa5.6 } - 2 \, \spa3.6\spb4.5 \biggr)  \cr
& \hskip 0.3 cm 
 + \biggl( { \spab3.{(1+4)}.5 \over \spb5.6 }
       - { \spa3.4\spb1.4\spa2.6 \over \spab2.{(5+6)}.1 } \biggr)
    { \spb4.5 \, \d12 - 2\,\spbb4.{36}.5 \over \spab2.{(5+6)}.1 \Delta_3 } \cr
& \hskip 0.3 cm 
+ 4 \, { \spab3.4.5 \spab6.{(1+3)}.4 + \spab6.3.4 \spab3.{(2+4)}.5  
         \over \spab2.{(5+6)}.1 \Delta_3 }   \cr
& \hskip 0.3 cm 
+ 2 \, { \d12 \over \spab2.{(5+6)}.1 \Delta_3 } 
    \biggl( { \spb4.5 \spab3.{(2+4)}.5 \over \spb5.6 }
         - { \spa3.6 \spab6.{(1+3)}.4 \over \spa5.6 } \biggr)
%%%%% end WW: logcoeff
\,, \cr
} \label{logcoeff}
\eeq
and the three-mass triangle coefficient $\Ta$ is given by
\beqn
\label{TaWW} \Ta & = & 
%%%%% begin WW Ta: Ta
% scalar:
  {3\over2} { s_{12} \, \d12 \, (t_{134}-t_{234})
               \spab6.{(1+2)}.5 \spab3.{(1+2)}.4
         \over \spab2.{(5+6)}.1 \Delta_3^2 }  \\
&-& {1\over2} { (3\,s_{12} + 2\, t_{134}) \spab6.{(1+2)}.5 \spab3.{(1+2)}.4
         \over \spab2.{(5+6)}.1 \Delta_3 } \nn \\
&+& { t_{134} \over {\spab2.{(5+6)}.1}^2 \Delta_3 }
\biggl[ \spb1.4\spa2.6 \bigl( \spab3.6.5 \, \d56 - \spab3.4.5 \, \d34 \bigr)
 \nn \\
&&  \qquad \qquad \qquad \quad
-\ \spb1.5\spa2.3 \bigl( \spab6.5.4 \, \d56 - \spab6.3.4 \, \d34 \bigr) \biggr]
\nn \\
&+& { \spa3.6\spb4.5 \, s_{12} \, t_{134} \over \spab2.{(5+6)}.1 \Delta_3 }  
- { \spa3.4\spb5.6 {\spab6.{(1+2)}.4}^2 \over \spab2.{(5+6)}.1 \Delta_3 } 
+ 2\, { \spa1.6\spb2.4 \bigl( \spab6.5.4 \, \d56 - \spab6.3.4 \, \d34 \bigr)
    \over \spb3.4\spa5.6 \Delta_3 }   \nn \\
% gf:
&+& 2\, { \spab6.{(2+5)}.4 \over \spab2.{(5+6)}.1 \Delta_3 }
   \biggl[ 
   { \spab6.5.2\spab2.1.4 \, \d56 - \spab6.2.1\spab1.3.4 \, \d34
   + \spab6.{(2+5)}.4 \, s_{12} \, \d12 
       \over \spb3.4\spa5.6 }\nn \\
&& \qquad \qquad \qquad \quad 
   + \ 2 \, \spab3.{(2+6)}.5 \, s_{12} \biggr] \nn \\
& -& { \spb1.4\spa2.6 \spab3.{(2+6)}.5 \over {\spab2.{(5+6)}.1}^2 }
+ 2 \, { \spb1.5\spa2.3 \spab6.{(2+5)}.4 \over {\spab2.{(5+6)}.1}^2 }
- { \spb1.4\spa2.6 \spab6.{(2+5)}.4 \, \d12 
     \over \spb3.4\spa5.6 {\spab2.{(5+6)}.1}^2 } \nn \\
&+& {1\over2} {1\over\spab2.{(5+6)}.1} \biggl[
  3 \, { \spab6.2.4\spab6.1.4 \over \spb3.4\spa5.6 }
  + { \spab3.2.5\spab3.1.5 \over \spa3.4\spb5.6 } 
  + { \spb1.4\spa1.6\spb4.5 \over \spb3.4 }
  - { \spb2.4\spa2.6\spa3.6 \over \spa5.6 } \nn \\
&&  \qquad \qquad \qquad \quad + \ { \spa2.3\spb2.5\spa3.6 \over \spa3.4 }
  - { \spa1.3\spb1.5\spb4.5 \over \spb5.6 }
  + 4 \, \spa3.6\spb4.5 \biggr] \nn \\
&+& {1\over2}{1\over\spab1.{(5+6)}.2} \biggr[ 
  { \spa1.6^2\spb2.4^2 \over \spb3.4\spa5.6 }
- { \spa1.3^2\spb2.5^2 \over \spa3.4\spb5.6 } \biggr] 
- {1\over2} { \spb1.4^2\spa2.6^2 (t_{134} \, \d12 + 2\,s_{34}\,s_{56})
    \over \spb3.4\spa5.6 {\spab2.{(5+6)}.1}^3 }
%%%%% end WW Ta: Ta
  \,. \nn
\eeqn
% The above was converted back into maple and checked against
% the WW maple file, 3/2/98.  -L.D.
%
For the reader's convenience we recall the definitions of the functions
appearing in eq.~(\ref{Fa}),
\beqn
\label{defsFun}
\Ll_0(r) &\equiv& \frac{\ln(r)}{1 - r}, \qquad
\Ll_1(r) \equiv \frac{\Ll_0(r) + 1}{1 - r} ,
   \nn \\
\Ls_{-1}(r_1,r_2) &\equiv& \li (1-r_1) + \li (1-r_2) + 
   \ln(r_1) \ln(r_2) - \frac{\pi^2}{6} \,, \nn \\
\Lsnew^{2 {\rm m} h}_{-1}(s,t,m_1^2,m_2^2) &\equiv& 
  - \li \left(1 - \frac{m_1^2}{t}\right)
  - \li \left(1 - \frac{m_2^2}{t}\right)
   \label{IntFunctions}  \\
  &-& \frac{1}{2} \ln^2 \left(\frac{-s}{-t}\right)
   +  \frac{1}{2} \ln \left(\frac{-s}{-m_1^2}\right) 
                 \ln \left(\frac{-s}{-m_2^2}\right)  \,, \nn \\
\eeqn
where the dilogarithm is
\beq
\li (x) = - \int_0^x dy \frac{\ln(1-y)}{y} \,.
\eeq
The analytic structure of the three-mass-triangle integral 
$I_3^{3 {\rm m}}$ is rather complicated in the general case 
(see Appendix A of ref.~\cite{BDKW4q}). However, for the production of a
vector boson pair $V_1 V_2$ we have  
$\Delta_3 = s_{12} (s_{12} -2 M_1^2 -2 M_2^2) + (M_1^2-M_2^2)^2 > 0$, 
i.e. we are in region (1b) of \cite{BDKW4q} and the integral is
simply given by
\beqn
\label{threemassint}
 I_3^{3 {\rm m}}(s_{12},s_{34},s_{56}) &=& - \frac{1}{\sqrt{\Delta_3}}
  {\rm Re} \bigg[ 2 ( \li(-\rho x) + \li(-\rho y)) 
                 + \ln(\rho x) \ln(\rho y)  \\
    & & \qquad +\ \ln\left(\frac{y}{x}\right)  
                 \ln\left(\frac{1 + \rho y}{1 + \rho x} \right)
       + \frac{\pi^2}{3}  \bigg] , \nn
\eeqn
where
\beq
x = \frac{s_{12}}{s_{56}}, \quad y = \frac{s_{34}}{s_{56}}, \quad
\rho = \frac{2 s_{56}}{\delta_{56} + \sqrt{\Delta_3}} \,.
\eeq

The rational-function terms in \eqn{Fa} have been rearranged some,
so they no longer look exactly like the permutation of the addition of
$\delta p$ in \eqn{deltap} to the rational-function terms of eq.~(3.20) 
of \cite{BDKW4q}.

%%%%%%%%%%%%%%%%%%%%%%%%%%%%%%%%

\subsection{Real (Bremsstrahlung) primitive amplitudes}

A full next-to-leading order calculation of vector-boson pair production
requires also the tree-level amplitudes with an additional gluon radiated
from the quarks. The corresponding primitive amplitudes are easily calculated.
For the case of a positive-helicity gluon with momentum $k_7$ 
they are given by:
\beqn
\label{bremsampA}
A_7^{\tree,a} &=& 
%%%%% begin WW:  Asevena
i \, {\spa1.3 \over \spa1.7 s_{34}\,s_{56}\,t_{134}} \left[ 
   {\spa1.3\spb3.4\spb2.5\spab6.{(2+5)}.7 \over t_{256}}
 + {\spab6.{(1+3)}.4 \spab1.{(2+7)}.5 \over \spa7.2} \right]
%%%%% end WW:  Asevena
  \\
\label{bremsampB}
A_7^{\tree,b} &=& 
%%%%% begin WW:  Asevenb
{i \over \spa1.7\spa7.2 s_{34}\,s_{56}\,t_{127}}  \Bigl[
  - \spa3.6\spb4.5 \spaa1.{(5+6)(2+7)}.1 \\
&& \qquad \qquad 
  + \ \spa1.3 \spab1.{(2+7)}.4 \spab6.{(3+4)}.5 
  - \ \spa1.6 \spab1.{(2+7)}.5 \spab3.{(5+6)}.4 \Bigr] \nn
%%%%% end WW:  Asevenb
\,. 
\eeqn
% The above was converted back into maple and checked against
% the WW maple file, 3/2/98.  -L.D.
The case of a negative-helicity gluon is given simply by applying
the operation $-\flip{1}$ to the positive-helicity case, 
where $\flip{1}$ is defined in eq.~(\ref{FlipSymmetry}). 
These amplitudes do have the appropriate limits when the gluon momentum
$k_7$ becomes soft, or becomes collinear with a quark momentum, 
$k_1$ or $k_2$.

%%%%%%%%%%%%%%%%%%%%%%%%%%%%%%%%

\section{Dressing with electroweak couplings for \newline
  $W^+ W^-,\ Z \, Z,\ W^\pm Z $    \label{dressing} } 
\setcounter{equation}{0}

This section is devoted to the construction of the (squared) amplitudes
from the primitive amplitudes given in the previous section.  We will
discuss each process in turn. The formulas are given for the tree-level
case with only four leptons in the final state.  With some trivial
modifications which will be discussed at the end of this section, the same
formulas can be used for the one-loop (squared) amplitudes and the
tree-level (squared) amplitudes with an additional gluon in the final
state.   

Concerning the labeling of the external particles:
The incoming antiquark always gets label 1, 
while the incoming quark always gets label 2.
The external gluon, if present, gets label 7.
The final-state lepton labelings correspond to the 
minimal modifications of the $WW$ case,
\beq
W^-\ +\ W^+\ \to\ (\ell_3 \, \bar\nu_4)\ +\ (\ellb^\prime_5 \, \nu^\prime_6),
\label{labelww}
\eeq
namely,
\beqn
\eqalign{
Z\ +\ Z\ &\to\ (\ell_3 \, \bar\ell_4)\ +\ (\ellb^\prime_5 \, \ell^\prime_6), 
\cr
W^-\ +\ Z\ &\to\ (\ell_3 \, \bar\nu_4)\ +\ (\ellb^\prime_5 \, \ell^\prime_6), 
\cr
}\label{labelzzwz}
\eeqn
and the $W^+Z$ case can be obtained by a CP transformation.

The results will be given in the unphysical configuration where
all particles are outgoing. The momenta have to satisfy 
$\sum_{i = 1}^6 p_i = 0 $. In this configuration the label 1 corresponds
to the outgoing quark; upon crossing to the physical region it becomes 
the label for the incoming antiquark.  Recall that the helicity of the 
quarks will change sign under the crossing operation. 

%%%%%%%%%%%%%%

\subsection{ $ \qb q \to W^- W^+ \to 
             \ell \nub_\ell \, \ellb' \nu_{\ell'} $ }

First consider up-quark annihilation into a $W^+ W^-$ pair.
The leptons (anti-leptons) have to be left-handed (right-handed).
If the (outgoing) up-quark is left-handed, the tree amplitude is
\beq
\eqalign{
\cA^\tree(u^L_1,\ub^R_2;\ell_3,\bar{\nu}_4;
         \bar{\ell}^\prime_5,\nu^\prime_6)
&= \left( {e^2\over\sstw} \right)^2 \, \delta_{i_1}^{\ib_2} 
   {s_{34}\over s_{34}-M_W^2+i \Gamma_W M_W}
   {s_{56}\over s_{56}-M_W^2+i \Gamma_W M_W} \cr
&\quad\times  
  \left[ A^{\tree,a}(1,2,3,4,5,6)
       + C_{L,u} A^{\tree,b}(1,2,3,4,5,6) \right] \,, \cr
}\label{uubarlefttree}
\eeq
where
\beq
C_{L,\{ {u\atop d} \}} = \pm 2 Q \sstw 
+ {s_{12} ( 1 \mp 2 Q \sstw ) \over s_{12} - M_Z^2} \,,
\label{CleftWWZ}
\eeq
$\theta_W$ is the weak mixing angle,
and we take $Q = 2/3$ and the upper sign in \eqn{CleftWWZ} for the up quark. 
The color structure of the amplitude is simply given by
$\delta_{i_1}^{\ib_2}$, where $i_1, \ib_2$ are the color labels of the
(anti-)quarks. Note that as $s_{12} \to \infty$,  
$C_{L,\{ {u\atop d} \}} \to 1$, and there is a high-energy cancellation
between $A^{\tree,a}$ and $A^{\tree,b}$:  
$A^{\tree,a} \sim - A^{\tree,b}$ as $s_{12} \to \infty$.
If the up-quark is right-handed, the tree amplitude is
\beq
\eqalign{
\cA^\tree(u^R_1,\ub^L_2;\ell_3,\bar{\nu}_4;
         \bar{\ell}^\prime_5,\nu^\prime_6)
&= \left( {e^2\over\sstw} \right)^2 \, \delta_{i_1}^{\ib_2} 
   {s_{34}\over s_{34}-M_W^2+i \Gamma_W M_W}
   {s_{56}\over s_{56}-M_W^2+i \Gamma_W M_W} \cr
&\quad\times  
  C_{R,u} A^{\tree,b}(2,1,3,4,5,6) \,, \cr
}\label{uubarrighttree}
\eeq
where
\beq
C_{R,\{ {u\atop d} \}} = \pm 2 Q \sstw \left[ 1 
- {s_{12}\over s_{12} - M_Z^2 } \right] \,. 
\label{CrightWWZ}
\eeq
Since $C_{R,\{ {u\atop d} \}} \to 0$ as as $s_{12} \to \infty$,
the high-energy cancellation is simpler in this case.
The finite width $\Gamma_Z$ of the
$Z$ boson can safely be neglected in eqs.~(\ref{CleftWWZ}) and
(\ref{CrightWWZ}) because we have $s_{12} > 4 M_W^2$. On the other hand, 
it is of course crucial to keep the $i\Gamma_W M_W$ terms in the
propagators of the $W$ bosons.

The corresponding tree amplitudes for down-quark annihilation are 
\beq
\eqalign{
\cA^\tree(d^L_1,\db^R_2;\ell_3,\bar{\nu}_4;
         \bar{\ell}^\prime_5,\nu^\prime_6)
&= \left( {e^2\over\sstw} \right)^2 \, \delta_{i_1}^{\ib_2} 
   {s_{34}\over s_{34}-M_W^2+i \Gamma_W M_W}
   {s_{56}\over s_{56}-M_W^2+i \Gamma_W M_W} \cr
&\quad\times  
  \left[ A^{\tree,a}(1,2,6,5,4,3)
       + C_{L,d} A^{\tree,b}(1,2,6,5,4,3) \right] \cr
}\label{ddbarlefttree}
\eeq
and 
\beq
\eqalign{
\cA^\tree(d^R_1,\db^L_2;\ell_3,\bar{\nu}_4;
         \bar{\ell}^\prime_5,\nu^\prime_6)
&= \left( {e^2\over\sstw} \right)^2 \, \delta_{i_1}^{\ib_2} 
   {s_{34}\over s_{34}-M_W^2+i \Gamma_W M_W}
   {s_{56}\over s_{56}-M_W^2+i \Gamma_W M_W} \cr
&\quad\times  
  \, C_{R,d} A^{\tree,b}(2,1,6,5,4,3) \,, \cr
}\label{ddbarrighttree}
\eeq
where we take $Q = -1/3$ and the lower sign in eqs.~(\ref{CleftWWZ}) and
(\ref{CrightWWZ}) for the down quark.
Note that because of Cabibbo-Kobayashi-Maskawa mixing and the large
mass of the top quark, we have not done the $t$-channel exchange
of the top quark correctly.  Fortunately this error is proportional
to the tiny quantity $|V_{td}|^2$ (or for $d\bar{s}$ annihilation,
to $V_{td} V^*_{ts}$ times a suppression factor for the strange sea quark
distribution in the proton).

We can now construct the differential cross-section in the narrow-width 
approximation.  We normalize the squared amplitudes $\cM^\tree$ such that 
the integral over center-of-mass angles for the lepton-pair decay products 
of both vector bosons only has to be multiplied 
by the two-body phase-space factor, in order to obtain the total partonic 
vector-boson pair production cross-section multiplied by the leptonic
branching ratios,
\beq
\label{dsig2}
d\sigma_2(q\qb \to V_1 V_2) \times B_\ell(V_1) \, B_{\ell^\prime}(V_2)
 = d\Phi_2 \int d^2\Omega_1 \int d^2\Omega_2 \ \cM^\tree \,.
\eeq
The two-body phase-space factor is given by
\beq
\label{phase-space2}
d\Phi_2 = \frac{\beta}{16\pi}\ d\cos\theta_{CM} \ ,
\eeq
where 
\beq
\beta = \frac{\sqrt{(s_{12}-(M_1+M_2)^2) (s_{12}-(M_1-M_2)^2)}}{s_{12}}
\eeq
and
\beq
\label{intOmega}
\int d^2\Omega_1 = 
\int_0^\pi d\theta_1 \sin\theta_1 \int_0^{2\pi} d\phi_1 \ .   
\eeq
Here $(\theta_1,\phi_1)$ are the angles of one of the decay leptons, 
measured in that vector boson $V_1$'s center-of-mass frame. For equal
masses $M_1=M_2$, $\beta$ is the velocity of $V_1$ and $V_2$ in the
$q\qb$ center-of-mass frame.

We obtain for the squared Born amplitude for unpolarized $\ub u \to WW$, 
\beqn
\lefteqn{{\cal M}^\tree(u_1, \ub_2 ;\ell_3,\bar{\nu}_4;
         \bar{\ell}^\prime_5,\nu^\prime_6) = B_\ell^2 (W)  
 \left( {e^2\over\sstw} \right)^2 \left({3\over4\pi}\right)^2 
  { M_W^4 \over 8 \, s_{12} \, N_c } }   \label{uuWWcross} \\
&&
\times\Biggl\{
   \left\vert A^{\tree,a}(1,2,3,4,5,6) 
    + C_{L,u} A^{\tree,b}(1,2,3,4,5,6) \right\vert^2 
    + \left\vert C_{R,u} A^{\tree,b}(2,1,3,4,5,6) \right\vert^2
          \Biggr\} \nn\  , 
\eeqn
while for unpolarized $\db d \to WW$, we obtain
\beqn
\lefteqn{ {\cal M}^\tree(d_1, \db_2;\ell_3,\bar{\nu}_4;
         \bar{\ell}^\prime_5,\nu^\prime_6 ) = B_\ell^2 (W)  
 \left( {e^2\over\sstw} \right)^2 \left({3\over4\pi}\right)^2 
  { M_W^4 \over 8 \, s_{12} \, N_c } } \label{ddWWcross} \\
&&
\times\Biggl\{
   \left\vert A^{\tree,a}(1,2,6,5,4,3) 
    + C_{L,d} A^{\tree,b}(1,2,6,5,4,3) \right\vert^2 
    + \left\vert C_{R,d} A^{\tree,b}(2,1,6,5,4,3) \right\vert^2
          \Biggr\} \nn \ ,  
\eeqn
with $N_c=3$ for QCD.
The corresponding expressions for longitudinally-polarized scattering 
can be obtained simply by dropping the ``unwanted'' terms in 
eqs. (\ref{uuWWcross}) and (\ref{ddWWcross}), and adjusting the 
normalization appropriately.

%%%%%%%%%%%%%%%%%

\subsection{ $ \qb q \to Z Z  \to \ell \ellb \, \ellb' \ell' $ }

In the $\qb q \to ZZ$ process, only the symmetric combination appears,
\beq
A^{\tree,s} = A^{\tree,a}(1,2,3,4,5,6) + A^{\tree,a}(1,2,6,5,4,3).
\label{atreesym}
\eeq
(This sum includes the `exchange' graphs just as in $A_6^\subl$;
hence the simpler form found for this quantity in \cite{BDK2q2g}
could be used here, once diagram (b) is removed and the
appropriate permutation is applied.)
In the $ZZ$ case, there are more non-vanishing helicity configurations.
We refrain from writing down all the individual helicity amplitudes in
this case, and just present the  differential cross-section
formulas for the various helicity configurations, normalized as discussed
above: 
\beqn
\label{qqZZcross} 
\cM^\tree(q_1^L,\qb_2^R; \ell_3^-, \ellb_4^+; \ellb_5^+, \ell_6^-)
 & = & p_{\rm ZZ}  \ 
   v_{L,q}^4  v_{L,e}^4 \, | A^{\tree,s}(1,2,3,4,5,6) |^2 \,, \nn  \\
\cM^\tree(q_1^L,\qb_2^R; \ell_3^+, \ellb_4^-; \ellb_5^+, \ell_6^-)
 & = & p_{\rm ZZ} \
   v_{L,q}^4  v_{L,e}^2  v_{R,e}^2 \, | A^{\tree,s}(1,2,4,3,5,6) |^2 \,,   \\
\cM^\tree(q_1^L,\qb_2^R; \ell_3^-, \ellb_4^+; \ellb_5^-, \ell_6^+) 
 & = & p_{\rm ZZ}  \ 
   v_{L,q}^4  v_{L,e}^2  v_{R,e}^2 \, | A^{\tree,s}(1,2,3,4,6,5) |^2 \,, \nn \\
\cM^\tree(q_1^L,\qb_2^R; \ell_3^+, \ellb_4^-; \ellb_5^-, \ell_6^+)
 & = & p_{\rm ZZ} \
   v_{L,q}^4  v_{R,e}^4  \, | A^{\tree,s}(1,2,4,3,6,5) |^2 \,, \nn
\eeqn
where we defined
\beq
\label{ZZprefactor}
p_{\rm ZZ} =  B_\ell^2 (Z) \, e^4 \left({3\over4\pi}\right)^2 
  { M_Z^4 \over 4 \, s_{12} \, N_c \, (v_{L,e}^2 + v_{R,e}^2)^2 }  \,.
\eeq
The formulas for $\cM^\tree(q_1^R,\qb_2^L;\ldots)$ can be obtained by
replacing $v_{L,q}^4$ by $v_{R,q}^4$ and interchanging the labels 1 and
2 in \eqn{qqZZcross}.
The unpolarized cross-section, normalized as in \eqn{dsig2}, is
given by the sum over all these $\cM^\tree$'s.
In the above, the left- and right-handed couplings to the $Z$ are
\beq
\eqalign{
v_{L,e} &= { -1 + 2\sin^2 \theta_W \over \sin 2 \theta_W } \,, 
\hskip 2.3 cm 
v_{R,e}  = { 2 \sin^2 \theta_W \over  \sin 2 \theta_W } \,,  
\cr
v_{L,q} &= { \pm 1 - 2 Q\sin^2 \theta_W \over  \sin 2 \theta_W } \,,
\hskip 1.9 cm 
v_{R,q} = -{ 2 Q \sin^2 \theta_W \over \sin 2 \theta_W }  \,, \cr
}\label{vLRdef}
\eeq
where $Q$ is the charge of quark $q$ in units of $e$, 
and the two signs in $v_{L,q}$ correspond to up $(+)$ and 
down $(-)$ type quarks.

%%%%%%%%%%%

\subsection{ $ \ub d \to W^- Z  \to  \ell \nub_\ell \, \ellb' \ell' $ }

Finally, for $WZ$ production the down-quark and the lepton $\ell$ have to
be left-handed. The lepton $\ell'$ which couples to the $Z$ can have either
polarization. If it is left-handed, the cross-section is proportional to
$v_{L,e}^2$ while for a right-handed $\ell'$ it is proportional to
$v_{R,e}^2$. For the unpolarized cross-section we have
\beqn
\lefteqn{\hspace*{-2cm}
{\cal M}^\tree(u_1, \db_2;\ell_3, \nub_3; \ellb'_4, \ell'_5)
  =  B_\ell (W)  B_{\ell'} (Z)     
 \vert V_{ud} \vert^2 
 \left( { e^2 \over \sin\theta_W  }\right)^2
 \left({3\over4\pi}\right)^2 
  { M_W^2 M_Z^2 \over 4\, s_{12} \, N_c \, (v_{L,e}^2 + v_{R,e}^2) } } 
\nn  \\
&
\times\Bigg\{
   v_{L,e}^2 \bigg| 
   v_{L,d} A^{\tree,a}(1,2,3,4,5,6) + v_{L,u} A^{\tree,a}(1,2,6,5,4,3) \nn \\
& - \ \cot\theta_W \frac{ s_{12}}{ s_{12}-M_W^2 } A^{\tree,b}(1,2,3,4,5,6)
     \biggr|^2 \label{udWZcross} \\
& \quad + \  v_{R,e}^2 \bigg| 
   v_{L,d} A^{\tree,a}(1,2,3,4,6,5)
 + v_{L,u} A^{\tree,a}(1,2,5,6,4,3) \nn \\
& - \ \cot\theta_W  \frac{s_{12}}{ s_{12}-M_W^2 } A^{\tree,b}(1,2,3,4,6,5)
     \bigg|^2 \Bigg\} \,. \nn
\eeqn
The results for the polarized cross-section can easily be extracted from
eq.~(\ref{udWZcross}). 

%%%%%%%%%%%%%%

\subsection{Loop and Bremsstrahlung amplitudes \label{LoopBrem}}

The QCD loop corrections to all the amplitudes under consideration are
given by a simple substitution of loop primitive amplitudes for tree
primitive amplitudes,
\beq
\cA^\lloop =  g^2 \, {N_c^2-1\over N_c} \,  
   \cA^\tree \Big|_{ A^{\tree,\alpha} \to A^\alpha}\ ,\quad \alpha=a,b  \,,
\label{allloopamps}
\eeq
where $g$ is the strong coupling, and $N_c = 3$ for QCD.  
Thus, the next-to-leading order virtual
corrections to the above cross-sections  are obtained by replacing 
$ (A^{\tree,\alpha})^* A^{\tree,\alpha} $ in
eqs.~(\ref{uuWWcross}), (\ref{ddWWcross}), (\ref{qqZZcross}) and
(\ref{udWZcross}) by $2 \ {\rm Re} [ (A^{\tree,\alpha})^*  A^\alpha ] $
and multiplying by $g^2 {N_c^2-1\over N_c} = 8 \pi \as C_F$.

The leading high-energy behavior of the loop amplitude cancels in the
same way as the tree amplitude.  First note that none of the terms in
$F^a$ contains the factor ${1\over s_{34}s_{56}}$, which is present in
$A^{\tree,a}$.  Thus the high-energy limit is governed completely by
the $V$ pieces.  This ensures that the tree-level cancellation
continues to take place at the loop level.

We remark that any non-Standard Model modifications of the three-boson
vertex would only affect the second ($b$) class of diagrams.
The virtual QCD corrections can therefore be obtained in this case simply 
by multiplying the modified tree amplitude by the factor $V$ 
given in \eqn{Vab}. 

The squared amplitudes for the processes with an additional external gluon
can be obtained by a similar change. In this case one has to replace 
$A^{\tree,\alpha}$ and $(A^{\tree,\alpha})^*$ by $A^{\tree,\alpha}_7$ and 
$(A^{\tree,\alpha}_7)^*$ respectively and multiply again by 
$g^2 \frac{\Nc^2 - 1}{\Nc}$. Furthermore, $s_{12}$ has to be replaced by
$t_{127}$ in the expressions for $C_{L,\{ {u\atop d} \}}$ and 
$C_{R,\{ {u\atop d} \}}$, eqs.~(\ref{CleftWWZ}) and (\ref{CrightWWZ}), as
well as in eq.~(\ref{udWZcross}). The normalization is the same as long as 
the two-body phase-space factor $d\Phi_2$ in eq.~(\ref{phase-space2}) is
replaced by the three-body phase-space factor $d \Phi_3$.  
Because eqns.~(\ref{bremsampA}) and (\ref{bremsampB}) for 
$A^{\tree,\alpha}_7$ are crossing-symmetric, these substitutions are 
equally valid for processes where the gluon is in the final state,
$q\qb \to V_1 V_2 g$, or in the initial state, $gq \to V_1 V_2 q$ and
$g\qb \to V_1 V_2 \qb$, provided that the factor coming from the
spin-color average is changed accordingly; all three processes contribute
at next-to-leading order.

%%%%%%%%%%%%%%%

\section{Real photons in the final state \label{photons}}
\setcounter{equation}{0}

The cases with a single real photon, $W\gamma$ and $Z\gamma$,
can be handled as well, but the required primitive amplitudes are
different.  The one-loop primitive amplitude needed for the $Z\gamma$ case 
can be extracted from the existing subleading-in-color primitive amplitude
for $e^+e^- \to q\qb g$, as given in appendix IV of \cite{BDK2q2g}.
This helicity amplitude was first calculated in \cite{GieleGlover}.
In the $W\gamma$ case, just as for the $W^+W^-$ and $W^-Z$ cases above,
the $e^+e^- \to q\qb g$ amplitude has to be split further into two 
`fully primitive' pieces (which in this case are not related by a symmetry).

\subsection{Virtual primitive amplitudes for $q\qb \to V_1 \gamma$}

For $q\qb \to V_1 \gamma$ there is no `exchange' symmetry to relate
the contributions with the reversed ordering of $V_1$ and the photon on the
quark line.  On the other hand, under an on-shell gauge transformation for the
photon (i.e., substitution of its polarization vector by its momentum),
the diagrams where the photon is radiated from the quark line mix
with those where it is radiated from the $V_1 = W$ line.
Let the photon be leg 5, the $q\qb$ pair be $\{1,2\}$,
and the leptonic decay products of $V_1$ be $\{3,4\}$.
Then we again define two primitive amplitudes,
\beq
A_\gamma^\alpha(1^-,2^+,3^-,4^+,5^+), \qquad \alpha=a,b,
\label{GammaPrimAmpDef1}
\eeq
where $A_\gamma^a$ ($A_\gamma^b$) includes all ordered diagrams where 
the photon is on the leg 2 (leg 1) side of the diagram.
The negative-helicity photon cases are obtained by another
`flip' permutation,
\beqn
A_\gamma^a(1^-,2^+,3^-,4^+,5^-) 
&=& \flip{2} \Bigl[ A_\gamma^b(1^-,2^+,3^-,4^+,5^+) \Bigr] \,, \nn \\
A_\gamma^b(1^-,2^+,3^-,4^+,5^-) 
&=& \flip{2} \Bigl[ A_\gamma^a(1^-,2^+,3^-,4^+,5^+) \Bigr] \,, 
\label{NegHelGamma}
\eeqn
where
\beq
\flip{2}: \hskip 8 mm
1\leftrightarrow 2\,, \hskip 0.4 cm  
3\leftrightarrow 4\,, \hskip 0.4 cm  
\spa{a}.{b} \leftrightarrow \spb{a}.{b} \,. 
\label{Flip2Symmetry}
\eeq
Therefore we present only the positive-helicity photon case below.

The tree amplitudes are given by 
\beqn
A_\gamma^{\tree,a} &=& 
%%%%% begin WW:  Agamtreea
- i \, { \spa1.3^2\spb1.5 \over \spa3.4\spa2.5\,(s_{12}-s_{34}) } 
%%%%% end WW:  Agamtreea
\,, \label{gammatreea} \\
A_\gamma^{\tree,b} &=& 
%%%%% begin WW:  Agamtreeb
- i \, { \spa1.3^2\spb2.5 \over \spa3.4\spa1.5\,(s_{12}-s_{34}) } 
%%%%% end WW:  Agamtreeb
\,, \label{gammatreeb}
\eeqn
where $s_{34}$ should be set equal to $M_{V_1}^2$ in accordance with
the narrow-width approximation.

For the one-loop photon amplitudes we use the same 
decomposition (\ref{VFdecomp}) into a divergent part $V$, and
finite parts $F$ which are given by
\beqn
F_\gamma^a & = &
%%%%% begin WW: Fgama
  { \spa1.3^2 \over \spa1.5\spa3.4\spa2.5 }
     \Ls_{-1}\Bigl({-s_{25}\over-s_{34}},{-s_{12}\over-s_{34}}\Bigr) 
+ {1\over2} \, { \spa1.2\spa3.5\spb4.5\spb2.5 \over \spa2.5 }
      { \Ll_1\Bigl({-s_{34}\over-s_{12}}\Bigr) \over s_{12}^2 } \nn \\
&& \hskip0.2cm 
- {3\over2} \, { \spa1.3\spb4.5 \over \spa2.5 }
      { \Ll_0\Bigl({-s_{34}\over-s_{12}}\Bigr) \over s_{12} }
- {1\over2} { \spb2.4\spb4.5 \over \spb1.2\spb3.4\spa2.5 } 
%%%%% end WW: Fgama
\,, \label{Fgama}
\eeqn
\beqn
F_\gamma^b & = &
%%%%% begin WW: Fgamb
 { \spa1.2^2 \spa3.5^2 \over \spa3.4\spa1.5\spa2.5^3 }
      \Ls_{-1}\Bigl({-s_{12}\over-s_{34}},{-s_{15}\over-s_{34}}\Bigr) 
- {1\over2} { \spa2.3^2 \spb2.5 ^2 \spa1.5 \over \spa3.4\spa2.5 }
      { \Ll_1\Bigl({-s_{34}\over-s_{15}}\Bigr) \over s_{15}^2 } \nn \\
&& \hskip0.2cm 
+ { \spa2.3\spb2.5 \bigl(\spa1.2\spa3.5+\spa1.3\spa2.5\bigr)
      \over \spa3.4\spa2.5^2 }
      { \Ll_0\Bigl({-s_{34}\over-s_{15}}\Bigr) \over s_{15} } \nn \\
&& \hskip0.2cm 
- {1\over2} { \spa1.2^2\spa3.5^2\spb2.5 
      \over \spa3.4\spa1.5\spa2.5^2 } \Biggl[ 
  s_{15} \, { \Ll_1\Bigl({-s_{34}\over-s_{12}}\Bigr) \over s_{12}^2 } 
   + 3 \, { \Ll_0\Bigl({-s_{34}\over-s_{12}}\Bigr) \over s_{12} } \Biggr]
 \nn \\
&& \hskip0.2cm 
- {1\over2} { \spa2.3\spb2.5\spa3.5 \over \spb1.2\spa3.4\spa2.5^2 }
- {1\over2} { \spb1.4\spb2.5\spb4.5 \over \spb1.2\spb3.4\spb1.5\spa2.5 } 
%%%%% end WW: Fgamb
\,.  \label{Fgamb}
\eeqn
% The above was converted back into maple and checked against
% the Wgamma maple file, 3/2/98.  -L.D.
The integral functions appearing in the above equations are defined in
eqs.~(\ref{IntFunctions}).

These amplitudes can be extracted from the collinear limit of the
virtual-photon case, $q\qb\to W\gamma^*$, when the momenta of the 
virtual photon's decay leptons 5 and 6 become parallel.
The relevant primitive-amplitude combination is 
$A^a + {s_{12}\over s_{12}-s_{34}} A^b$ (see \eqn{udWZcross}).  In the
collinear limit, with $k_P \equiv k_5+k_6$, it becomes
\beq
A^a + {s_{12}\over s_{12}-s_{34}} A^b
\ \mathop{\longrightarrow}^{5 \parallel 6}\ 
{z\over\spb5.6} \, A_\gamma^a(1^-,2^+,3^-,4^+,P^+)
\ +\ {1-z \over \spa5.6} \, A_\gamma^a(1^-,2^+,3^-,4^+,P^-)
\,. \label{looplimit}
\eeq
Using the `flip' relation, we can read off the two independent
photon amplitudes.

The sum of $A_\gamma^a$ and $A_\gamma^b$ (which is all that is required
for the case  $q\qb \to Z\gamma$) reproduces, after relabelings, the
existing subleading-in-color primitive amplitude for $e^+e^- \to q\qb g$,  
as given in appendix IV of \cite{BDK2q2g}.

%%%%%%%%%%%%%%%%%%%%%%%%%%%%%%%%

\subsection{Real (Bremsstrahlung) primitive amplitudes}

The tree-level amplitudes with an additional gluon radiated
from the quarks, $q\qb \to V_1 \gamma g$, can similarly be obtained 
from the collinear limit of the $q\qb \to V_1 \gamma^* g$ case,
which is described by 
$A_7^{\tree,a} + {t_{127}\over t_{127}-s_{34}} A_7^{\tree,b}$.
Let the gluon label be 6.  In analogy to the definitions for the 
virtual case, $A_{6,\gamma}^{\tree,a}$ 
($A_{6\gamma}^{\tree,b}$) includes all ordered diagrams where 
the photon is on the leg 2 (leg 1) side of the diagram.
The results are
\beqn
&& \hspace*{-0.7cm} A_{6,\gamma}^{\tree,a}(1^-,2^+,3^-,4^+,5^+,6^+) = 
%%%%% begin WW: A6gamapp
i \, { \spa1.3^2 \spab2.{(3+4)}.5
   \over \spa3.4\spa2.5\spa1.6\spa6.2 \, (t_{126}-s_{34}) }
%%%%% end WW: A6gamapp
\ , \label{A6gamapp} \\
\lefteqn{A_{6,\gamma}^{\tree,a}(1^-,2^+,3^-,4^+,5^-,6^+) = } 
 \label{A6gamamp}\\
%%%%% begin WW: A6gamamp
&&  - i \, \Biggl[ 
  { \spa1.3\spb6.2\spab5.{(1+3)}.4 
    \over s_{34} \, \spb2.5\spa6.2 \, t_{134} }
+ { \spab1.{(2+6)}.4 
   \bigl( \spa3.4\spa1.5\spb4.2 + \spa3.5\spa1.6\spb6.2 \bigr)
    \over s_{34} \, \spb2.5\spa1.6\spa6.2 \, (t_{126}-s_{34}) }
\Biggr]
%%%%% end WW: A6gamamp
\ , \nn  \\
&& \hspace*{-0.7cm} A_{6,\gamma}^{\tree,b}(1^-,2^+,3^-,4^+,5^+,6^+) = 
%%%%% begin WW: A6gambpp
- i \, { \spa1.3^2\spab1.{(3+4)}.5
   \over \spa3.4\spa1.5\spa1.6\spa6.2 \, (t_{126}-s_{34}) }
%%%%% end WW: A6gambpp
\ , \label{A6gambpp} \\
\lefteqn{A_{6,\gamma}^{\tree,b}(1^-,2^+,3^-,4^+,5^-,6^+) = }
  \label{A6gambmp} \\
%%%%% begin WW: A6gambmp
&& i \, \Biggl[
 { \spa1.5\spb2.4\spab3.{(1+5)}.6 
   \over s_{34} \, \spb1.5\spa1.6 \, t_{234} } 
+ { \spab1.{(2+6)}.4 \bigl( \spaa5.{(2+6)4}.3 + \spa5.3 \, s_{26} \bigr)
   \over s_{34} \, \spb1.5\spa1.6\spa6.2 \, (t_{126}-s_{34}) }
\Biggr]
%%%%% end WW: A6gambmp
\ . \nn
\eeqn
% The above was converted back into maple and checked against
% the Wgamma maple file, 3/2/98.  -L.D.
The cases where the gluon has negative helicity are obtained by 
applying $+\flip{2}$ to the above amplitudes (which simultaneously reverses
the photon helicity).

%%%%%%%%%%%%

\subsection{Dressing with electroweak couplings \label{dressingPhot} }

As in the case of two massive vector bosons, we present the fully dressed
amplitudes and differential cross-sections at tree level.  The
substitutions required to obtain the one-loop (squared) amplitudes and
tree-level (squared) amplitudes with an additional gluon are exactly
the same as those described in section \ref{LoopBrem}.

\subsubsection{ $ \ub d \to W^- \gamma \to \ell \nub_\ell \, \gamma $ }

For the process $\ub d \to W^-\gamma$,
the lepton (anti-lepton) has to be left-handed (right-handed),
and the (outgoing) up-quark must also be left-handed. 
The tree amplitude is
\beq
\eqalign{
\cA^\tree(u_1,\db_2;\ell_3,\bar{\nu}_4;\gamma_5^\pm)
&= \sqrt{2} \left( {e^3\over\sstw} \right) V_{ud} \, \delta_{i_1}^{\ib_2} 
   {s_{34}\over s_{34}-M_W^2+i \Gamma_W M_W} \cr
&\quad\times  
  \left[ Q_2 \, A_\gamma^{\tree,a}(1,2,3,4,5^\pm)
         + Q_1 \, A_\gamma^{\tree,b}(1,2,3,4,5^\pm) 
  \right] \,, \cr
}\label{ubardgammatree}
\eeq
where $Q_1 = 2/3$ ($Q_2 = -1/3$) is the up (down) quark charge.
>From this amplitude we obtain the spin-summed squared amplitude,
\beqn
\lefteqn{
 \cM^{\tree}(u_1,\db_2; \ell_3, \bar{\nu}_4; \gamma_5)  =} \\
&& \nn   B_\ell (W) \, \vert V_{ud} \vert^2 
\left( \frac{e^2 }{ \sin \theta_W}  \right)^2
\left( \frac{3}{4 \pi}  \right)
\frac{M_W^2}{4\, s_{12}\, N_c } 
\sum_{\lambda = \pm}
\left| Q_2 A_\gamma^a(1,2,3,4,5^\lambda) +
       Q_1 A_\gamma^b(1,2,3,4,5^\lambda) \right|^2  .
\eeqn
The normalization of this cross-section is the same as in the $W^+ W^-$
case, \eqn{dsig2}, except that the integral $\int d^2\Omega_2$ and the 
factor $B_{\ell^\prime}(V_2)$ should be omitted.

%%%%%%%%%%%%%%%%%

\subsubsection{ $\qb q \to Z \gamma  \to \ell \ellb \, \gamma$ }

In the $\qb q \to Z\gamma$ process, only the symmetric combination appears,
\beq
A_\gamma^{\tree,s}(1,2,3,4,5^\pm) = 
A_\gamma^{\tree,a}(1,2,3,4,5^\pm) + A_\gamma^{\tree,b}(1,2,3,4,5^\pm).
\label{agamtreesym}
\eeq
We again refrain from writing down all the individual helicity amplitudes 
here, and just present the  differential cross-section
formulas for the various helicity configurations, normalized as discussed
above: 
\beqn
\label{qqZgamcross} 
\cM^\tree(q_1^L,\qb_2^R; \ell_3^-, \ellb_4^+; \gamma_5^\pm) 
 & = & p_{Z \gamma}  \ 
   Q^2 v_{L,q}^2  v_{L,e}^2 
  \, | A_\gamma^{\tree,s}(1,2,3,4,5^\pm) |^2 \,, \nn  \\
\cM^\tree(q_1^L,\qb_2^R; \ell_3^+, \ellb_4^-; \gamma_5^\pm)
 & = &  p_{Z \gamma} \
   Q^2 v_{L,q}^2 v_{R,e}^2 \, | A_\gamma^{\tree,s}(1,2,4,3,5^\pm) |^2 \,,  
\eeqn
where we defined
\beq
\label{gamZZprefactor}
p_{Z \gamma} =  B_\ell (Z) \, e^4 \left({3\over4\pi}\right) 
  { M_Z^2 \over 2 \, s_{12} \, N_c \, (v_{L,e}^2 + v_{R,e}^2) }  \,.
\eeq
The formulas for $\cM^\tree(q_1^R,\qb_2^L;\ldots)$ can be obtained by
replacing $v_{L,q}^2$ by $v_{R,q}^2$ and interchanging the labels 1 and
2 in \eqn{qqZgamcross}.

%%%%%%%%%%%%%%%%

\section{Concluding remarks \label{conclusion}}
\setcounter{equation}{0}

We have presented all helicity amplitudes which are needed for a complete
computation of the next-to-leading order QCD corrections to the production
of a $Z\,Z, \ W^+ W^-, \ W^\pm Z, \ W^\pm \gamma$ or $Z \gamma $ pair at
hadron colliders, where the spin correlations are fully taken into
account.  The subsequent decay of each massive vector boson into a lepton
pair is included in the narrow-width approximation.  

The above cross-sections have been ``integrated'' over the 
lepton decay angles, after which they reproduce the previously published
analytic formulae for the virtual corrections of
refs.~\cite{ZZit,WZit,WWit,STN} to 30 digits accuracy. 
Such a high accuracy can be achieved because the true integration can be
replaced by $6\times6=36$ numerical evaluations, in which each lepton is
emitted along three orthogonal axes (both positive and negative directions)
in the corresponding vector-boson center-of-mass frame. As a check of the
bremsstrahlung amplitudes we compared in the same way the real
contribution in the $W^+W^-$ case to ref.~\cite{WWit} and found full
agreement.  

Of course the new information provided by the above cross-sections is
not the total cross-section, but rather the correlations between the 
lepton decay angles (or lab-frame momenta).  There are at least two
different ways to access this information at next-to-leading order:
\par\noindent
(1) Construct a general purpose Monte Carlo program directly in
terms of the lepton momenta.
\par\noindent
(2) Compute the elements of the density matrix 
$D_{i_1,i_2;j_1,j_2}(s,\ldots)$.  
These are the amplitude interferences for the production of a vector 
boson $V_1$ with helicity $i_1$ or $j_1 = -1,0,+1$ along its direction 
of motion (in the $q\qb \to V_1 V_2$ frame, for example), and $V_2$ with 
helicity $i_2$ or $j_2$.  The complete ${\cal O}(\alpha_s)$ density matrix
for polarized vector boson production can easily be computed using the
results presented in this paper. One simply has to carry out the
integration over the angles of the lepton pair, except that one now 
weights the numerical evaluations with an additional projection operator,
$\exp[ i (-i_1 + j_1) \phi_1 + i (-i_2 + j_2) \phi_2 ]$,
where $\phi_{1,2}$ is the azimuthal angle of a decay lepton with respect
to the direction of motion of $V_{1,2}$.  One can then fold the computed
production density matrix with the decay density matrices.
\par\noindent
A possible advantage of the second approach is that it is straightforward 
to include the additional QCD corrections that are present for
hadronically decaying vector bosons, such as those computed in 
\cite{AbLampe} for the $W$ boson.  

For this latter application, to processes such as 
$q\qb \to W^+W^- \to q^\prime\qb^{\prime\prime} \, \ell \nub_\ell$ 
(and neglecting interferences with non-resonant processes), 
the required amplitudes may be obtained from
those described in this paper by appropriate modifications of the coupling
constant factors.  (At ${\cal O}(\alpha_s)$, gluon exchange between the
initial and final quark lines gives a vanishing correction.)  Because
quark final states are phenomenologically somewhat less relevant, these
amplitudes have not been explicitly presented here.  Similarly, one can 
easily extend these results to cover the cases where the lepton pair
coming from an on-shell $Z$ boson is replaced by a Drell-Yan pair of 
arbitrary invariant mass from an intermediate virtual photon plus $Z$.

The amplitudes presented here can easily be implemented in a Monte Carlo
program. This would extend previous results \cite{OhnSpin} in that the
spin correlations of the virtual part would also be fully included. 
Furthermore, the simplicity of the amplitudes should aid in the 
next-to-leading-order study of the effects of non-standard 
triple-gauge-boson couplings. 

\bigskip
\bigskip

{\large \bf Acknowledgments}
\bigskip

We would like to thank S. Frixione for useful discussions and D.A. Kosower
for use of his tex-to-maple converter. A.S. would like to thank the Theory
Group of ETH Z\"urich, where part of this work has been done, for its
hospitality. Z.K. is grateful to the CERN Theory Group for its hospitality.

%%%%%%%%%%%%%%

\def\np#1#2#3  {{\it Nucl. Phys. }{\bf #1} (19#3) #2}
\def\nc#1#2#3  {{\it Nuovo. Cim. }{\bf #1} (19#3) #2}
\def\pl#1#2#3  {{\it Phys. Lett. }{\bf #1} (19#3) #2}
\def\pr#1#2#3  {{\it Phys. Rev. }{\bf #1} (19#3) #2}
\def\prl#1#2#3  {{\it Phys. Rev. Lett. }{\bf #1} (19#3) #2}
\def\prep#1#2#3 {{\it Phys. Rep. }{\bf #1} (19#3) #2}
\def\zp#1#2#3  {{\it Zeit. Phys. }{\bf #1} (19#3) #2}
\def\rmp#1#2#3  {{\it Rev. Mod. Phys. }{\bf #1} (19#3) #2}

\end{document}